# Quantile-Quantile Methodology – Detailed Results

Douglas M Hawkins

School of Statistics

University of Minnesota

dhawkins@umn.edu

## Overview

The quantile-quantile (QQ) relationship connects the ordered values of a sample (ordinate) with the scores corresponding to the distribution the data are thought to follow (abscissa). If this distribution is the normal, then the scores are expected normal scores, or more-easily-computed surrogates. A linear relationship supports the appropriateness of the assumed distribution. If the assumed distribution is location-scale, such as the normal, then the location and scale parameters can be estimated from the intercept and slope of a regression line fitted to the pairs. If the distribution has a third parameter, for example shape, that parameter can be estimated as the value optimizing the QQ linearity. For example power transformations to normality can be fitted by picking the power leading to the most linear QQ relationship.

The QQ relationship can also be extended to censored data, and can be robustified to accommodate a few extreme values.

This research was motivated by finding reference intervals ("normal ranges") for analytes, in particular those seen in clinical chemistry – typically two-sided 95% intervals.  It extends immediately to the parallel problems in environmetrics and the social sciences.  The main focus of the results reported is the normal distribution.  Appendix A addresses easily-computed approximations to the expected normal scores, and finds one particular choice appropriate on the grounds that it gives near-perfect statistical efficiency for the least-squares intercept/slope estimators of the underlying true mean and standard deviation for samples of size 120, the default in traditional nonparametric reference interval studies.

An attraction of parametric analyses is that they can provide good results with smaller samples than are needed for nonparametric.  At the other extreme, good estimates of remote quantiles may need much larger samples,  Appendix B extended the results of Appendix A to both bigger and smaller samples, and shows that the intercept and slope of the least squares line fitted to the QQ relationship continues to give estimates practically as good as the theoretically optimal sample mean and standard deviation.

The QQ relationship extends to censored data: one blanks out the censored values and their scores and fits the line to the uncensored data.  Appendix C quantifies the effect of the censoring on the statistical properties of the resulting slope and intercept, and also on the default 95% reference limit that can be computed from the estimated mean and standard deviation.  The effect is quantified by the statistical efficiency of the censored sample estimates, and an effective sample size computed from the statistical efficiency.

Censoring is involuntary – instruments do not report out-of-range values – but deliberately censoring some small number of the highest and the lowest data leads to a methodology that is robust to a few wayward, possibly invalid, samples. We call this process "winsorizing" the QQ relationship because of its conceptual similarity with the winsorized mean used in robust estimation. Winsorizing loses some statistical efficiency; this is quantified in Appendix D.

Turning to the location-scale-shape setting, the Box-Cox power transformation is widely used to model non-normal data. This can be done within the QQ framework by making the Box-Cox transformation for each trial power $\lambda$ in a range, finding the correlation coefficient QQr of the resulting QQ relationship, and picking the $\lambda$ value maximizing this correlation. This approach, an alternative to the more conventional pseudolikehood turns out to have good statistical efficiency, as shown in Appendix E.

The correlation coefficient QQr of the QQ relationship can be used as a formal test of normality. Appendix F develops transformations to take QQr into a putative N(0,1) test statistic. Different transformations are needed depending on whether the QQ relationship is on the original or the Box-Cox transformed scale and whether it is or is not winsorized. These are discussed in Appendix F. A transformation for censored data is also presented.

Appendix G gives some evidence that the test for normality, while not quite as powerful as the Shapiro-Wilk, is not far behind.

Appendix H illustrates fitting a location-scale distribution with an additional parameter – the *t* distribution whose degrees of freedom $\nu$ are fitted to the data. This can be done in essentially the same way as fitting a Box-Cox distribution – a line search over $\nu$ with a QQ plot using appropriate *t* scores for each $\nu$ and finding the maximum QQr. Other one-parameter transformations to normality such as the Manly and the generalized log can be fitted with minor modifications to the Box-Cox procedure.

One attraction is that these methodologies are easy to program -- even in a spreadsheet – and can be implemented by lab scientists without specialist statistical skills or software.

## Appendix A. The normal scores

The abscissae of the QQ plot are normal scores – the expected values of the order statistics of a standard normal sample. There is no simple closed-form expression for them, so it is conventional to use approximations. One family of approximations is that of Blom (1958):

$$Z_i = \Phi^{-1}\left[\frac{i-\beta}{n+1-\alpha-\beta}\right] \qquad (A1)$$

where $\Phi^{-1}$ is the inverse of the standard normal cumulative distribution function, $n$ is the sample size, $Z_i$ is the approximation to the $i^{th}$ expected normal score, and $\alpha, \beta$ are constants to be chosen.

Harter (1961) and Hyndman and Fan (1996) discussed the choice of these constants. While any two values between 0 and 1 could in principle be used for $\alpha$ and $\beta$, for normal quantile plots (and indeed plots for any symmetric distribution) one would expect point symmetry in the scores: $Z_i = -Z_{n+1-i}$. Requiring this leads to

$$\Phi^{-1}\left[\frac{i-\beta}{n+1-\alpha-\beta}\right] = -\Phi^{-1}\left[\frac{n+1-i-\beta}{n+1-\alpha-\beta}\right],$$

$$\left[\frac{i-\beta}{n+1-\alpha-\beta}\right] = 1 - \left[\frac{n+1-i-\beta}{n+1-\alpha-\beta}\right], \qquad (A2)$$

which, after some simplification, gives

$$\alpha = \beta$$

The most widely-used choices are $\alpha=\beta=0.5$, $\alpha=\beta=0$ which correspond to the normal scores

$\Phi^{-1}\{(i-0.5)/n\}$ (the so-called Hazen scores) and

$\Phi^{-1}\{i/(n+1)\}$ (the Weibull scores).

One way to pick $\alpha=\beta$ is to consider the QQ plot regression as an estimator of $\mu$ and $\sigma$, whose optimal estimators are the sample mean and standard deviation. As the intercept of the QQ regression is the sample mean, this leads to picking the $\alpha=\beta$ value that maximizes the statistical efficiency of the estimate of $\sigma$. To explore this, 100,000 random normal samples of size 120 (the default in reference range studies) were generated, and the standard deviation and the QQ plot slope estimates $s$ of $\sigma$ calculated. Figure A1 shows a profile plot of the efficiency as measured by mean squared error as a function of this common value of $\alpha,\beta$. The dashed vertical line shows the empirical maximum at $\alpha=\beta=0.47$, and the dotted line shows the value at $\alpha=\beta=0.5$. As this plot shows, there is no perceptible loss of performance in using 0.5 leading to the Hazen score $\Phi^{-1}(i-0.5)/n$, rather than the empirical maximum. Blom's recommended (but not widely-used) suggestion $\alpha=\beta=0.375$, is also close to optimal. There is however a large loss in performance in going much further from 0.5. In particular, the Weibull score $i/(n+1)$ is poor.

This calculation leads to our recommendation of the approximate normal scores $\Phi^{-1}\{(i-0.5)/n$

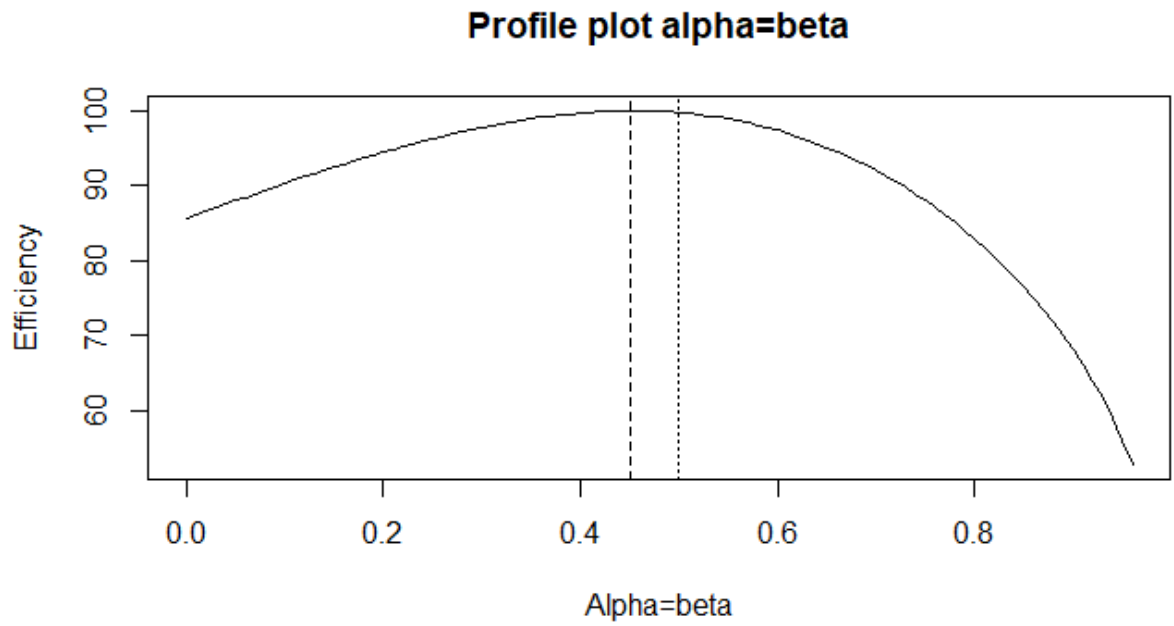

Figure A1. Efficiency of *s* as a function of $\alpha,\beta$

## Appendix B Efficiency of the QQ estimator

Expanding the simulation of Appendix A, 10,000 samples of sizes 30, 60, 120 and 240 were drawn from the standard normal distribution, the QQ plot constructed using the $\Phi^{-1}(i-0.5)/n$ formula for the normal scores and fitting the ordinary least squares (OLS) regression.

Figure B1 shows the two estimators in the first 100 samples; they appear to be all but identical, particularly at the larger sample sizes.

Table B1 shows the mean and standard deviation of the two for each sample size. The true standard deviation is 1, so both estimators are biased. The bias in $s$ could be corrected using the $c_4$ correction (Cureton [3]), but the biases of both estimators are so small this seems unnecessary.

The root mean square error RMSE, $\sqrt{(\text{bias}^2 + \text{sd}^2)}$, summarizes the typical distance of the estimator from the true value, incorporating both bias and random variability. The final column of Table B1, the relative efficiency, is the square of the ratio of the RMSE of $s$ to that of the QQ slope. As Figure B1 suggests, and as this column shows, the performance of the QQ slope as an estimator of $\sigma$ is, for practical purposes, the same as that of the sample standard deviation $s$. The standard error of the sample standard deviation is approximately $s/\sqrt{(2n)}$ and that of the sample mean is $s/\sqrt{n}$, and they are uncorrelated, properties that carry over to the slope and intercept of the QQ regression. The standard error of the estimated 97.5% point, $m+1.96s$ is then $\sigma\sqrt{[(1+1.96^2/2)/n]} = 1.71\sigma/\sqrt{n}$.

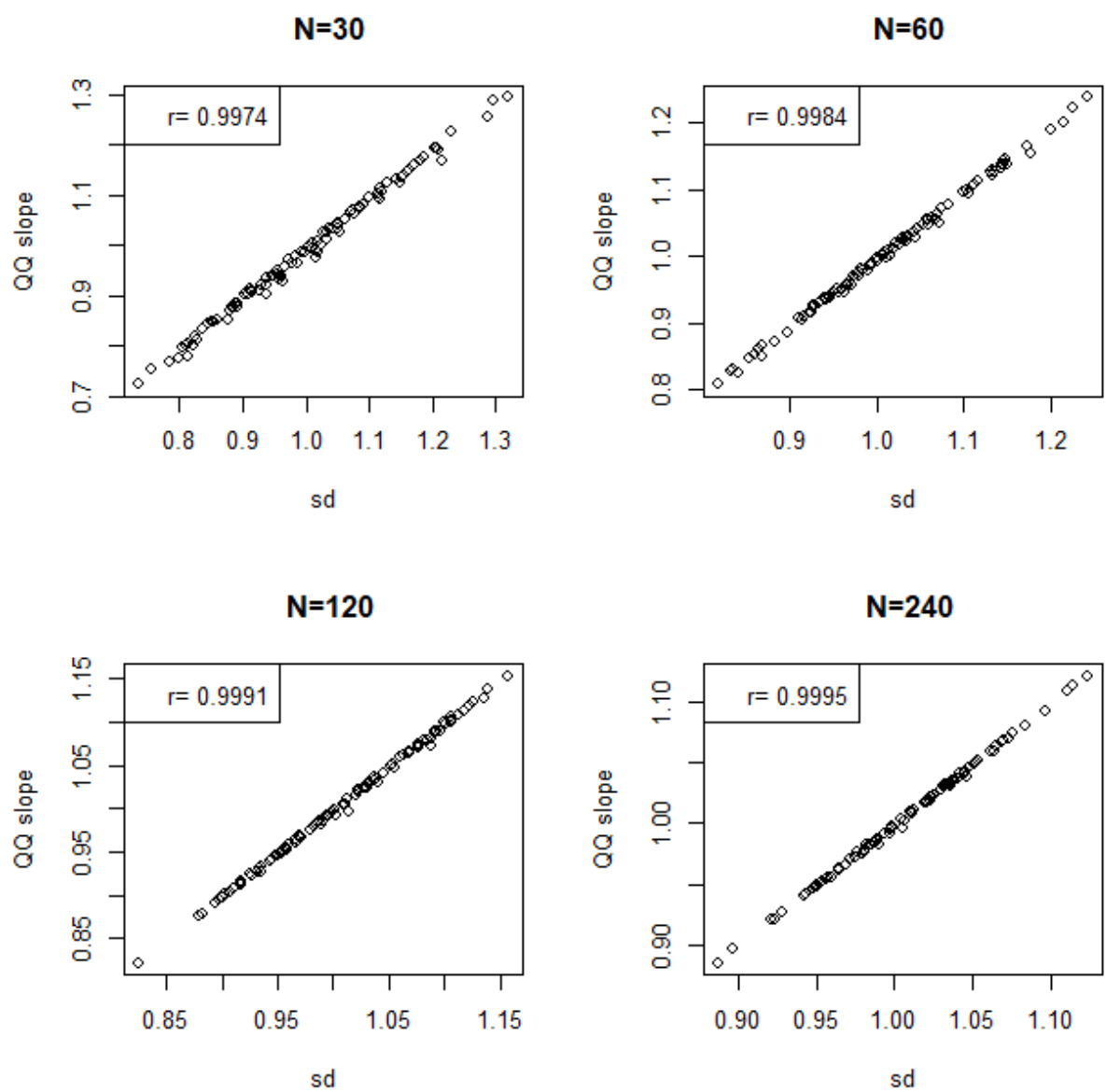

Figure B1. QQ slope vs sample sd

| N | mean | | sd | | RMSE | | Efficiency |
| --- | --- | --- | --- | --- | --- | --- | --- |
| | s | QQ slope | s | QQ slope | s | QQ slope | |
| 30 | 0.9919 | 0.9793 | 0.1305 | 0.1292 | 0.1308 | 0.1309 | 99.86 |
| 60 | 0.9958 | 0.9883 | 0.0920 | 0.0915 | 0.0921 | 0.0922 | 99.70 |
| 120 | 0.9982 | 0.9939 | 0.0646 | 0.0645 | 0.0647 | 0.0648 | 99.61 |
| 240 | 0.9991 | 0.9967 | 0.0455 | 0.0455 | 0.0455 | 0.0456 | 99.89 |

Table B1 Comparison of sample standard deviation and QQ slope as estimators of $\sigma$

# Appendix C  Effect of censoring

The intercept and slope of the QQ regression with censored data estimate the mean and standard deviation, like those of the full-sample case, but to get standard errors and confidence intervals requires assessment of the impact of the censoring.   This was done in a simulation of 100,000 samples of size 60, 80, … 200, 240, …400, 480…1080.  Each sample was subjected to 5, 10, … 50% left censoring and QQ plotted.  The statistical efficiency of each such censored sample was calculated as the ratio of the variance in the censored setting to the uncensored.

The reference limit problem anchoring this research aims by default for the 97.5% point of the distribution.  For normal data, this is $\mu+1.96\sigma$, whose natural estimate is $m+1.96s$.  Figure C1 provides an overview of inference on this estimated upper limit.  It suggests that the efficiency does not depend perceptibly on the sample size, but does depend on the amount of censoring.

Figure C2, averaged across all sample sizes, shows the efficiency of the estimates of the underlying mean, standard deviation, and reference limit.  Censoring has a large effect on the estimate of the standard deviation and less on the mean.  It is however much less deleterious for the estimate of the upper reference limit.  Intuition supports this; the censoring leaves the observations in the vicinity of the reference limit intact, retaining their information, shedding far-away observations.

Empirical models for the efficiency were fitted; these were

Mean            Efficiency = $1 - 1.5(k/n)^{1.7}$

sd  Efficiency = $(2.5 - 1.5f)^{-2}$

reference limit Efficiency = $(1.38 - 0.37f)^{-2}$

where *k* is the number of observations censored and *f* =1-*k*/*n* is the fraction uncensored. These fitted models are drawn in Figure C2 and capture the shape accurately.

It is convenient to turn these efficiencies into "effective" sample sizes. For example, if an estimator has a statistical efficiency of 75% relative to the fully efficient estimator, and is used in a sample of size *n*=120, then its precision is the same as that of a sample of size 0.75*120 = 90 using the fully efficient estimator. We say that the effective sample size for purposes of confidence interval calculation is 90.

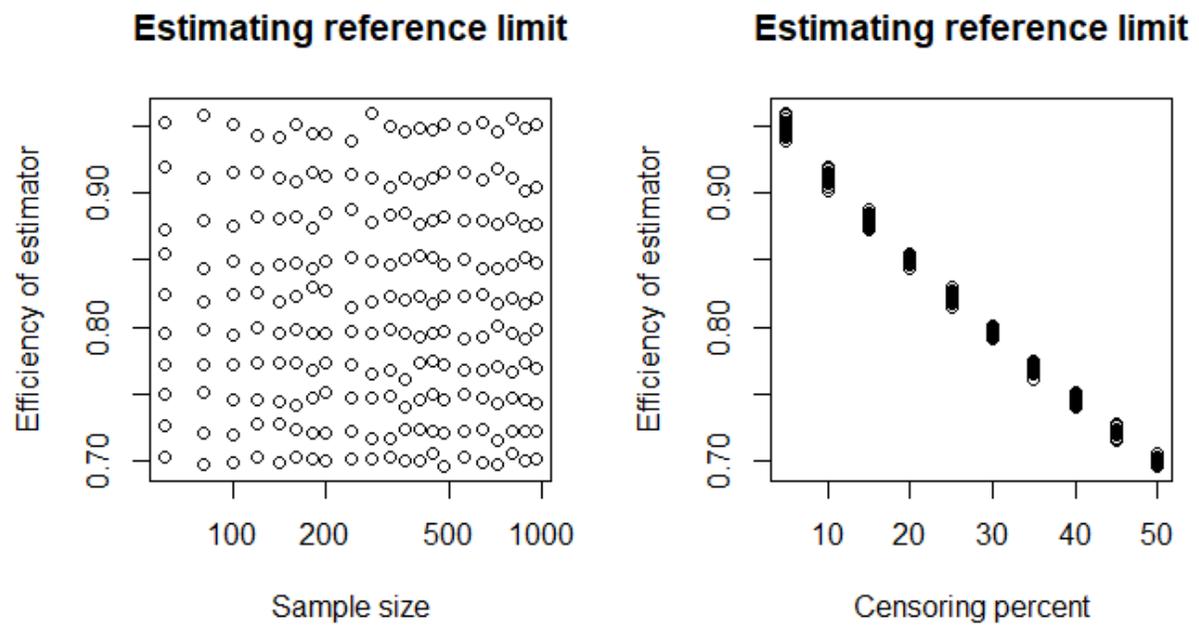

Figure C1. Efficiency of censored data by sample size and censoring

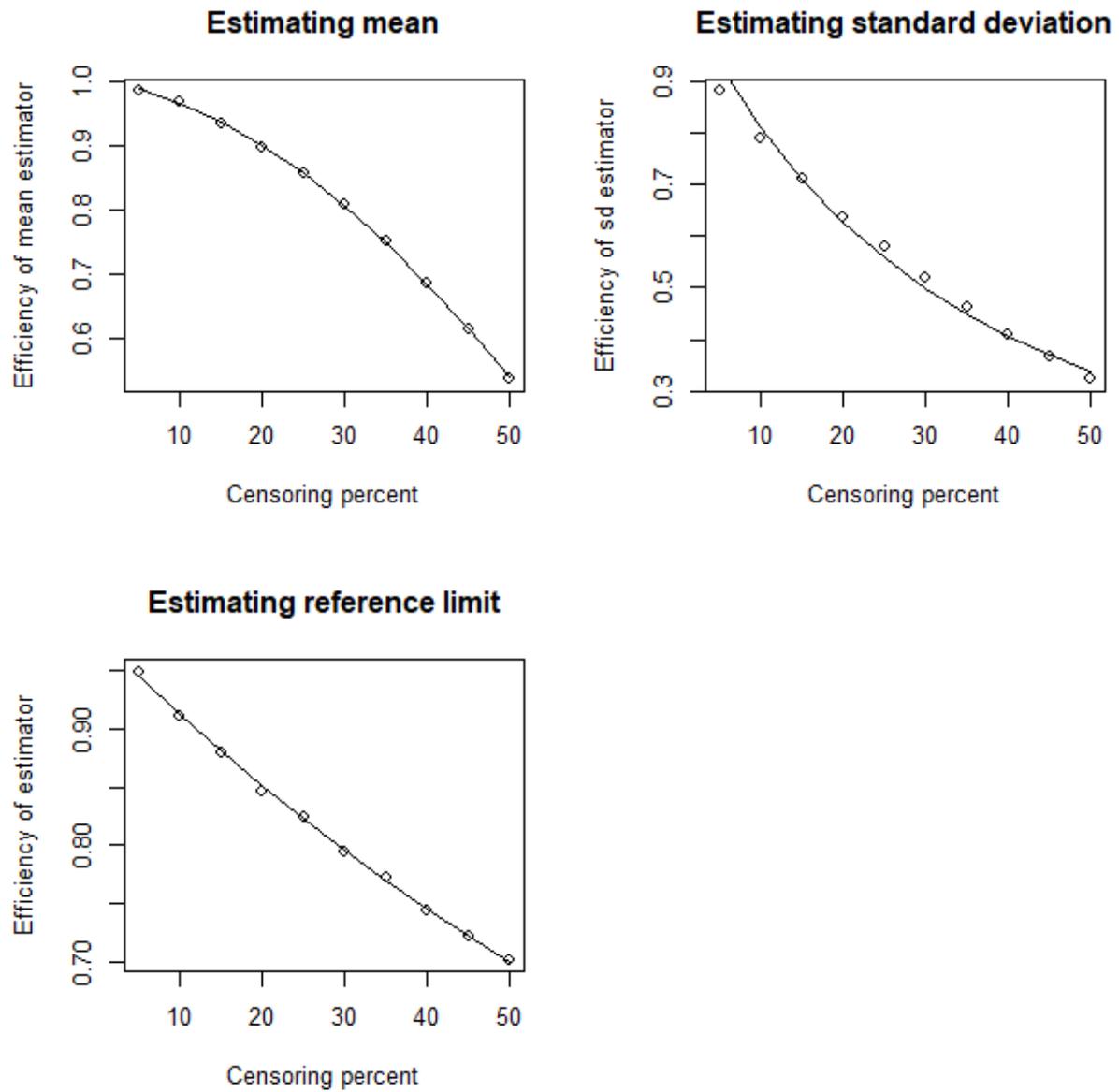

Figure C2. Efficiency of censored data for mean, sd and upper reference limit

## Appendix D. Winsorization

The winsorized QQ plot is constructed by omitting the *w* largest and smallest observations from the QQ regression. A simulation was run to explore the resulting estimators of mean, standard deviation and reference limit in the context of the 97.5% reference limit. Sample sizes of 80, 120, … 240 were used; 100,000 samples of each size generated from N(0,1), and 1, 2, 3, 4 and 5 observations winsorized on each side. The estimated 97.5 upper reference limit was calculated and its standard deviation found. Its standard deviation using the full sample is $1.71/\sqrt{n}$ per Appendix B. If we regard the winsorized sample as having some "effective" sample size $n_{eff}$, then the winsorized samples' estimated upper reference limit will have standard deviation $1.71/\sqrt{n_{eff}}$. From this, the observed standard deviations of the winsorized samples can be used to calculate the implied $n_{eff}$ and the sample size lost through winsorization $n - n_{eff}$

Table D1 summarizes these calculations of sample size lost to winsorization. The scattered negative entries for the mean reflect random variability. This table is described well by the model

Mean $\qquad n_{eff} = n$

SD $\qquad n_{eff} = n - 5w$

Reference limit $\qquad n_{eff} = n - 3.5w$

Winsorizing one more reading on each side, reducing the number of points in the QQ regression by 2, has little impact on the efficiency of the mean, but costs 5 readings for the SD and 3.5 for the 95% reference limits.

| Parameter | n | Winsorize | | | | |
|---|---|---|---|---|---|---|
| | | 1 | 2 | 3 | 4 | 5 |
| Mean | 80 | 0.4 | 0.3 | 1.6 | 3.3 | 3.2 |
| | 120 | 1.7 | 1.5 | 1.3 | 3.3 | 4.4 |
| | 160 | -0.9 | 2.0 | 2.3 | -1.8 | 1.0 |
| | 200 | -2.5 | 3.2 | 4.7 | 0.9 | 2.3 |
| | 240 | -0.7 | -0.9 | 0.1 | 2.2 | -2.7 |
| SD | 80 | 4.8 | 9.0 | 13.5 | 16.3 | 20.6 |
| | 120 | 5.0 | 10.2 | 12.9 | 19.1 | 21.6 |
| | 160 | 5.4 | 10.6 | 13.0 | 19.0 | 24.0 |
| | 200 | 8.3 | 9.5 | 13.2 | 21.9 | 25.4 |
| | 240 | 6.3 | 6.3 | 15.5 | 28.7 | 29.3 |
| 97.5% limit | 80 | 3.3 | 6.5 | 9.8 | 12.4 | 17.0 |
| | 120 | 3.6 | 7.2 | 10.3 | 15.9 | 16.7 |
| | 160 | 3.4 | 6.7 | 9.5 | 9.7 | 17.5 |
| | 200 | 5.6 | 9.8 | 11.6 | 16.4 | 16.6 |
| | 240 | 5.8 | 7.3 | 13.1 | 25.3 | 17.9 |

Table D1. Effective sample lost by winsorizing

# Appendix E. Efficiency of QQ estimator of $\lambda$

To use QQ rather than pseudolikelihood (PL) calls for evidence that it can recover a known true BC power with accuracy comparable with PL. The performance of the QQ estimator of the Box-Cox power was assessed by taking 10,000 samples of size 120 from a N(1, $0.25^2$) and raising them to powers $1/\lambda$ for $\lambda$ = -2, -1.75 … 2. For $\lambda$ = 0 the numbers were exponentially transformed. The Box-Cox model was then fitted to each sample by maximum QQr and by PL as in the original Box-Cox paper to see how closely each procedure reproduced the true power used to generate the data.

On an individual level, the average correlation between the QQ and PL estimates for the same $\lambda$ was 0.9957. Figure E1 shows the bias and sampling variability of the two methods. The optimal QQr had a smaller bias than did pseudolikelihood, but slightly larger variance.

The bias, sd and RMSE of the two methods are summarized in Table E1. The final column gives the RMSE efficiency of QQ relative to PL. This is high for all values of $\lambda$, averaging 92.2%.

| λ | Bias | | Sd | | RMSE | | Efficiency |
|---|---|---|---|---|---|---|---|
| | QQ | PL | QQ | PL | QQ | PL | |
| -2.0 | 0.014 | 0.073 | 0.592 | 0.564 | 0.592 | 0.568 | 92.0 |
| -1.5 | 0.011 | 0.056 | 0.440 | 0.419 | 0.440 | 0.422 | 92.0 |
| -1.0 | 0.012 | 0.042 | 0.293 | 0.279 | 0.293 | 0.282 | 92.6 |
| -0.5 | 0.005 | 0.020 | 0.146 | 0.139 | 0.146 | 0.140 | 92.1 |
| 0.0 | -0.004 | -0.004 | 0.315 | 0.306 | 0.315 | 0.306 | 94.4 |
| 0.5 | -0.003 | -0.017 | 0.148 | 0.140 | 0.148 | 0.141 | 91.4 |
| 1.0 | -0.009 | -0.039 | 0.294 | 0.279 | 0.294 | 0.282 | 91.8 |
| 1.5 | -0.013 | -0.058 | 0.451 | 0.429 | 0.452 | 0.433 | 91.8 |
| 2.0 | -0.012 | -0.071 | 0.585 | 0.556 | 0.585 | 0.561 | 91.7 |

Table E1. Efficiency of QQ estimate of BC power

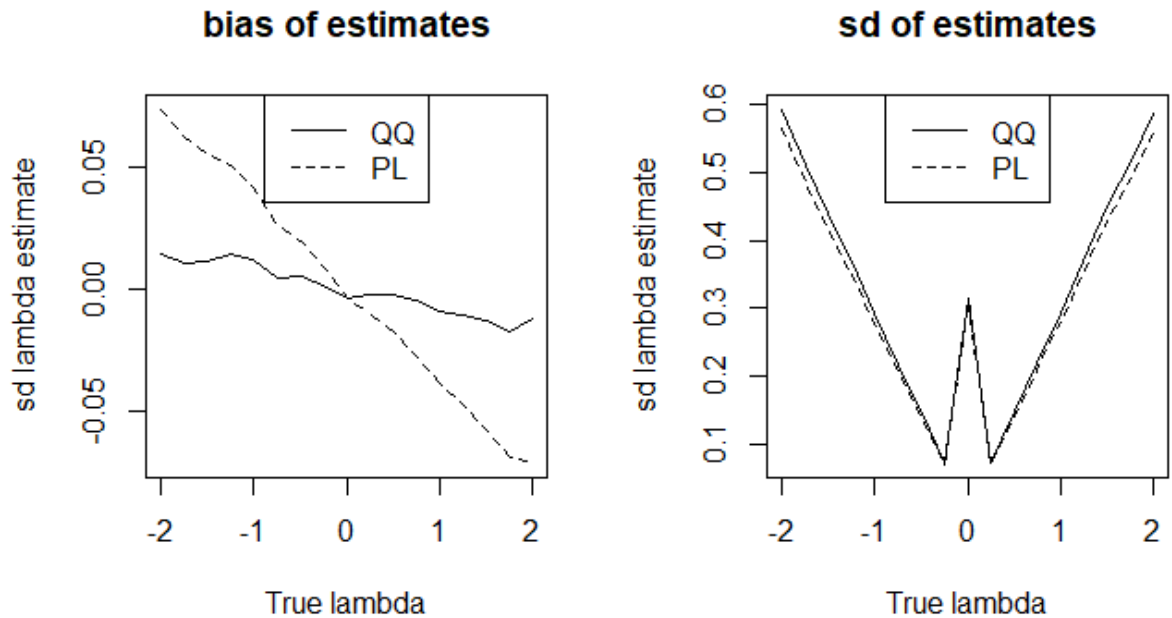

Figure E1. Estimation of Box-Cox power by QQ and pseudolikelihood

# Appendix F Testing normality by QQr

Turning the QQ correlation into a formal test of normality requires its statistical distribution. To address this, 13 sample sizes from 60 to 1080 in approximate geometric progression were used. For each sample size $n$, 100,000 samples were drawn from N(0,1). Four QQr correlations were found from each sample: one using the full sample, one winsorizing 2.5% in each tail; a BC analysis of the exp of the sample, and a winsorized BC analysis.

The sample size n=120 is most relevant for reference interval studies. For this sample size, the Box-Cox transformation bringing the QQr closest to normality was $\lambda=-0.1$, leading to

$$Z = \frac{(1-r)^{\lambda} - 1}{\lambda} \text{ with } \lambda = -0.1$$

The inverse, taking a $Z$ value back to the corresponding QQr is

$$r = 1 - (1 + \lambda Z)^{1/\lambda}$$

Figures F1 and F2 show the mean and the standard deviation of $Z$ as functions of $\ln(n+30)$ for each of the four procedures. Both plots are modeled well by straight lines, with the different procedures giving different lines. Fitting these lines, the mean and sd of the Z transform can be modeled as functions of $n$ by

Mean = A + B ln($n$+30)
sd   = D + E ln($n$+30)

Table F1 shows these regressions. The correlation from each procedure can be transformed to a random variable with mean zero and standard deviation 1 with the transformation

$$Z = \frac{Y - A - B\ln(n)}{D + E\ln(n)} = \frac{((1-r)^\lambda - 1)/\lambda - A - B\ln(n)}{D + E\ln(n)}$$

While Z has mean and sd close to 0 and 1, it is not a given that the BC transformation of 1-r will lead to a normal distribution, but this was verified using QQ plots of the transform of the correlations broken down by sample size and method. Illustrative plots are shown in Figure F3. As small values of r, suggesting non-normality, transform to large values of Z, the right tail of the distribution determines the adequacy of the test for normality and any breakaway from the line of identity in the left tail is immaterial.

The correlation can also be used with censored data, a counterpart to Royston's [24] suggestion of using the censored Shapiro-Wilk test. To explore this, the same suite of sample sizes was studied, fractions 0.05, 0.1, … 0.5 were left-censored and the QQr calculated and BC transformed using the same $\lambda$ value. The mean and standard deviations of the resulting transforms were then modeled with a bilinear equation

$$a + b[c + \ln(n+30)][d + f]$$

where $f$ is the fraction of the sample that is censored. The regressions leading to this fit are summarized in Table F2. The model was checked against the 10,000 samples used to calibrate each $n, f,$ combination and was found to match the standard normal closely.

|  | Term | Coeff | std_err | t_value | P_value |
|---|---|---|---|---|---|
| Mean full | (Intercept) | 1.992 | 0.03388 | 58.784 | 4.267e-15 |
|  | ln(n+30) | -1.802 | 0.005808 | -310.306 | 4.888e-23 |
|  | R_sqd | 0.9999 | Res_sd | 0.0164 |  |
| Winsor | (Intercept) | 3.12 | 0.03588 | 86.964 | 5.795e-17 |
|  | ln(n+30) | -2.115 | 0.00615 | -343.853 | 1.58e-23 |
|  | R_sqd | 0.9999 | Res_sd | 0.0173 |  |
| BC | (Intercept) | 1.405 | 0.04056 | 34.647 | 1.389e-12 |
|  | ln(n+30) | -1.782 | 0.006952 | -256.268 | 4.009e-22 |
|  | R_sqd | 0.9998 | Res_sd | 0.0196 |  |
| BC winsor | (Intercept) | 2.809 | 0.03562 | 78.875 | 1.694e-16 |
|  | ln(n+30) | -2.164 | 0.006105 | -354.443 | 1.132e-23 |
|  | R_sqd | 0.9999 | Res_sd | 0.0172 |  |
|  |  |  |  |  |  |
| sd full | (Intercept) | 0.6717 | 0.008341 | 80.526 | 1.349e-16 |
|  | ln(n+30) | 0.02561 | 0.00143 | 17.910 | 1.74e-09 |
|  | R_sqd | 0.9668 | Res_sd | 0.0040 |  |
| Winsor | (Intercept) | 0.4413 | 0.01323 | 33.357 | 2.1e-12 |
|  | ln(n+30) | 0.08462 | 0.002268 | 37.318 | 6.175e-13 |
|  | R_sqd | 0.9922 | Res_sd | 0.0064 |  |
| BC | (Intercept) | 0.5941 | 0.008564 | 69.369 | 6.938e-16 |
|  | ln(n+30) | 0.03245 | 0.001468 | 22.105 | 1.822e-10 |
|  | R_sqd | 0.9780 | Res_sd | 0.0041 |  |
| BC winsor | (Intercept) | 0.4288 | 0.01522 | 28.177 | 1.318e-11 |
|  | ln(n+30) | 0.07453 | 0.002608 | 28.572 | 1.133e-11 |
|  | R_sqd | 0.9867 | Res_sd | 0.0073 |  |

Table F1. Mean and SD of BC transform of QQr as a function of $n$

| Mean | Term | coeff | std_err | t_value | P_value |
|---|---|---|---|---|---|
| Original | (Intercept) | 2.256 | 0.04253 | 53.052 | 3.416e-83 |
|  | ln(n+30) | -1.923 | 0.00748 | -257.096 | 9.828e-162 |
|  | f | -0.7297 | 0.1371 | -5.323 | 5.047e-07 |
|  | f*ln(n+30) | 0.6353 | 0.02411 | 26.348 | 8.726e-51 |
|  | R_sqd | 0.9996 | Res_sd | 0.0295 |  |
| BC | (Intercept) | 1.796 | 0.05915 | 30.360 | 5.003e-57 |
|  | ln(n+30) | -1.937 | 0.0104 | -186.180 | 1.628e-145 |
|  | f | -1.331 | 0.1907 | -6.983 | 1.927e-10 |
|  | f*ln(n+30) | 0.7059 | 0.03354 | 21.050 | 2.003e-41 |
|  | R_sqd | 0.9992 | Res_sd | 0.0411 |  |
|  |  |  |  |  |  |
| sd | Term | Coeff | std_err | t_value | P_value |
| Original | (Intercept) | 0.598 | 0.01311 | 45.603 | 6.427e-76 |
|  | ln(n+30) | 0.05197 | 0.002307 | 22.531 | 3.512e-44 |
|  | f | 0.2236 | 0.04227 | 5.290 | 5.848e-07 |
|  | f*ln(n+30) | -0.01872 | 0.007435 | -2.518 | 0.01316 |
|  | R_sqd | 0.9529 | Res_sd | 0.0091 |  |
| BC | (Intercept) | 0.475 | 0.03021 | 15.725 | 1.565e-30 |
|  | ln(n+30) | 0.06489 | 0.005313 | 12.213 | 1.463e-22 |
|  | f | 0.3955 | 0.09736 | 4.062 | 8.873e-05 |
|  | f*ln(n+30) | -0.06081 | 0.01712 | -3.551 | 0.0005554 |
|  | R_sqd | 0.7803 | Res_sd | 0.0210 |  |

Table F2. Coefficients for model of censored data QQr

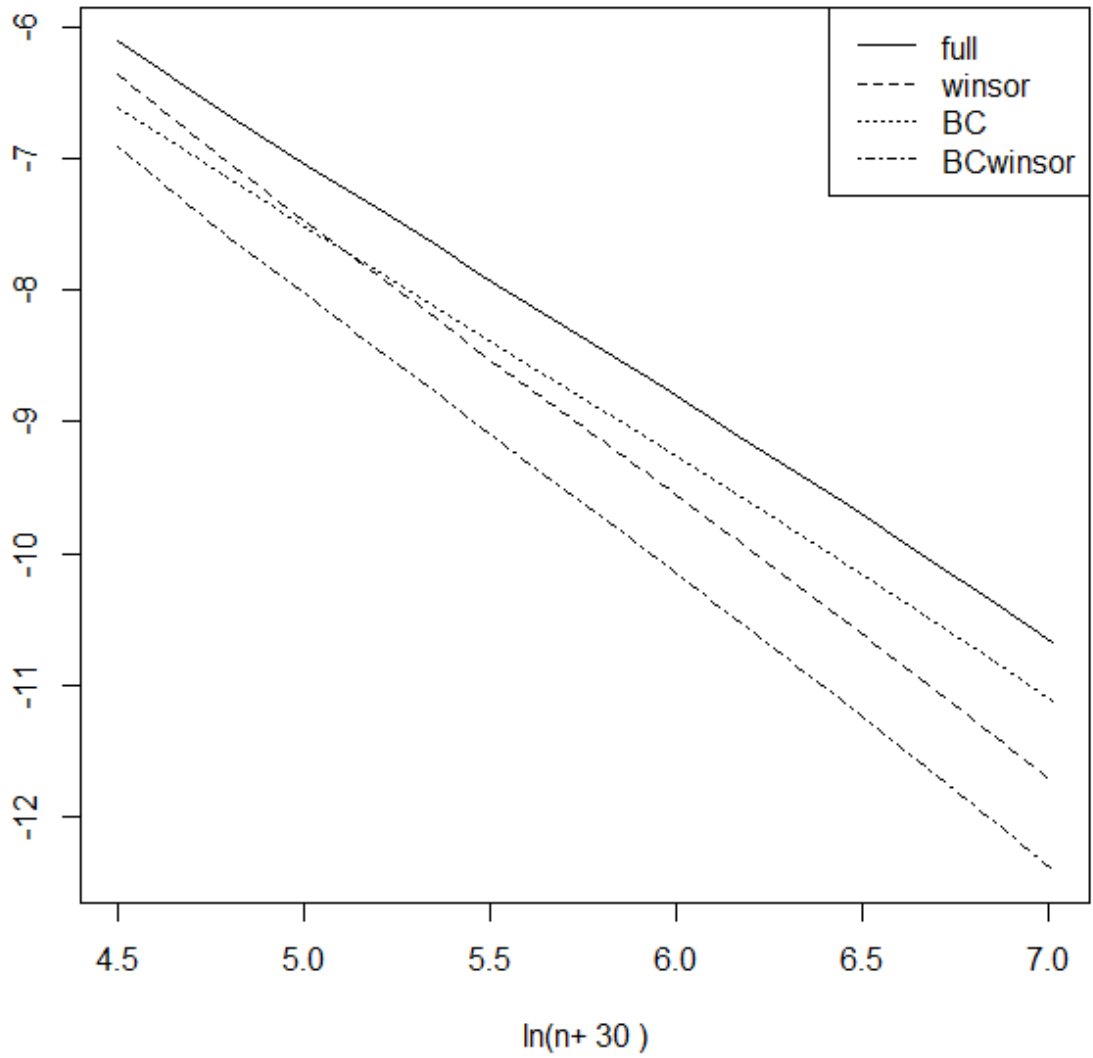

Figure F1. Mean of BC transform of QQr by n

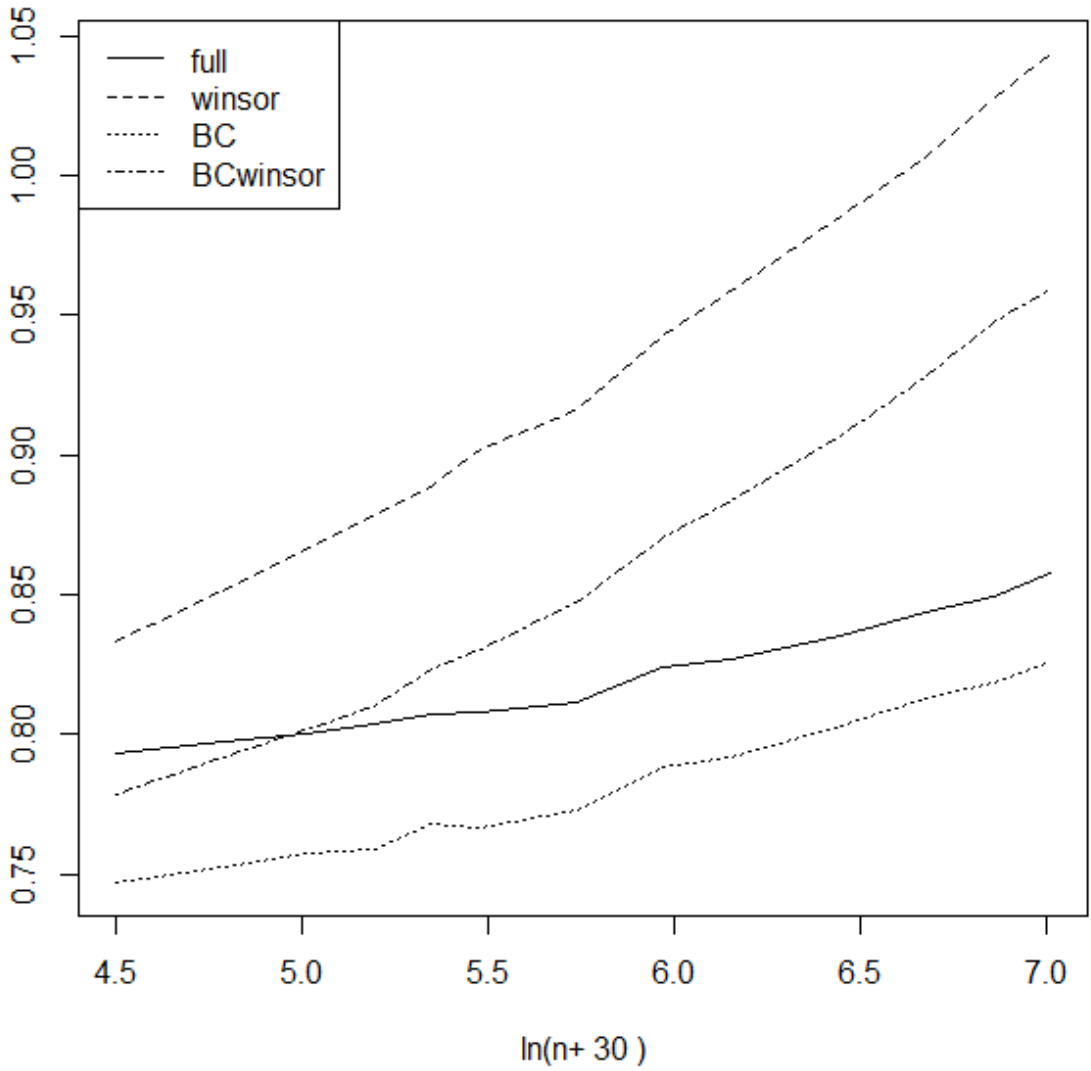

Figure F2. sd of BC transform of QQr by n

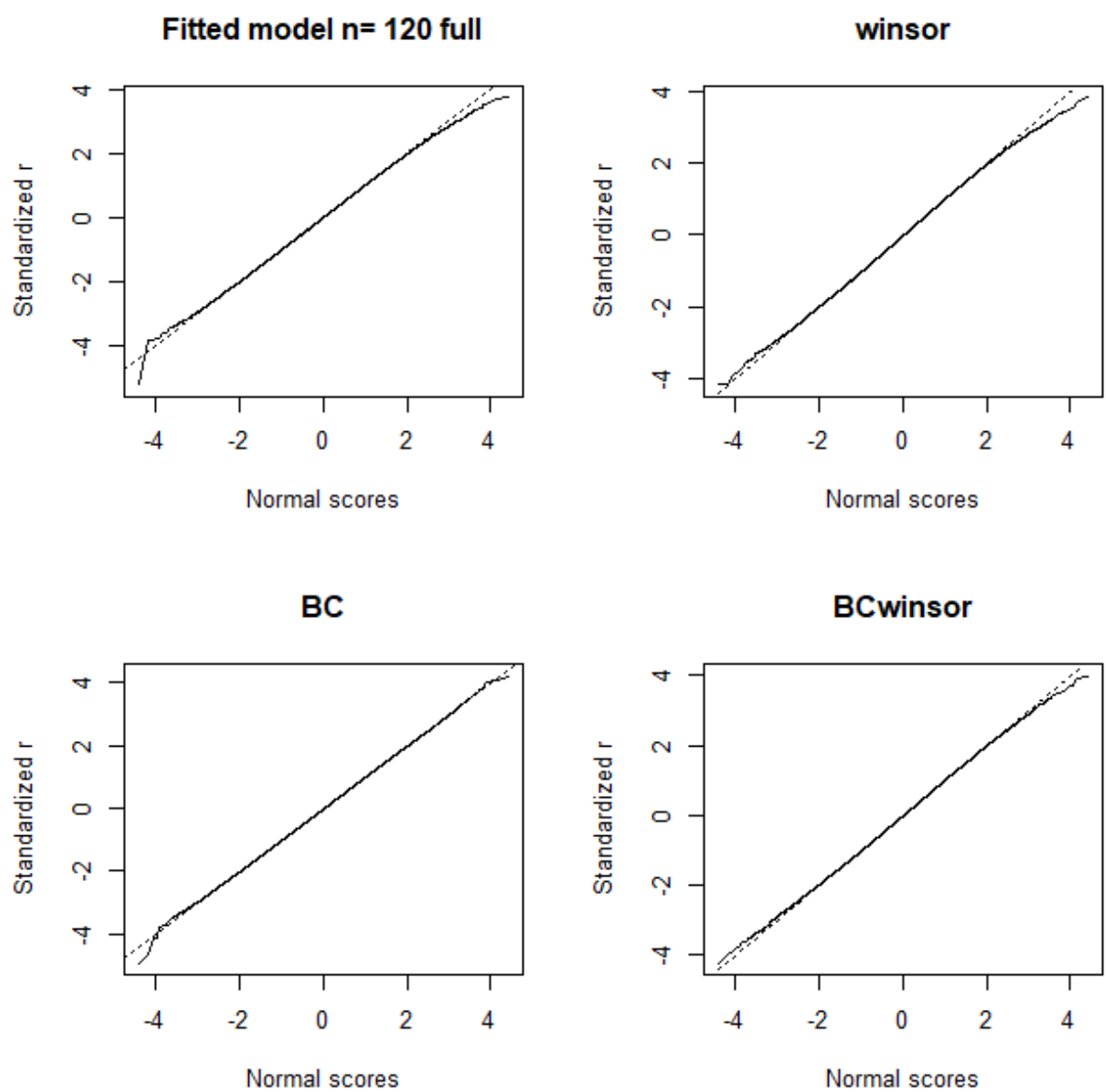

Figure F3. Distribution of BC transform of QQr for *n* =120

# Appendix G. Performance of QQr test for normality

The good performance of the normality test based on QQr for the right-skew data typically seen in clinical chemistry was verified using samples of size 120 of a normal(3,1) variable raised to a sequence of powers from 1 to 2, generating increasing right-skew non-normality. We call these successive departures from normality the "shift". For each such shift, 10,000 samples were generated and categorized by whether or not their P value, as assessed by QQr and SW, was below 0.05.

The results are displayed in Figure G1. While the Shapiro-Wilk test was more powerful than QQr, the difference in performance was small.

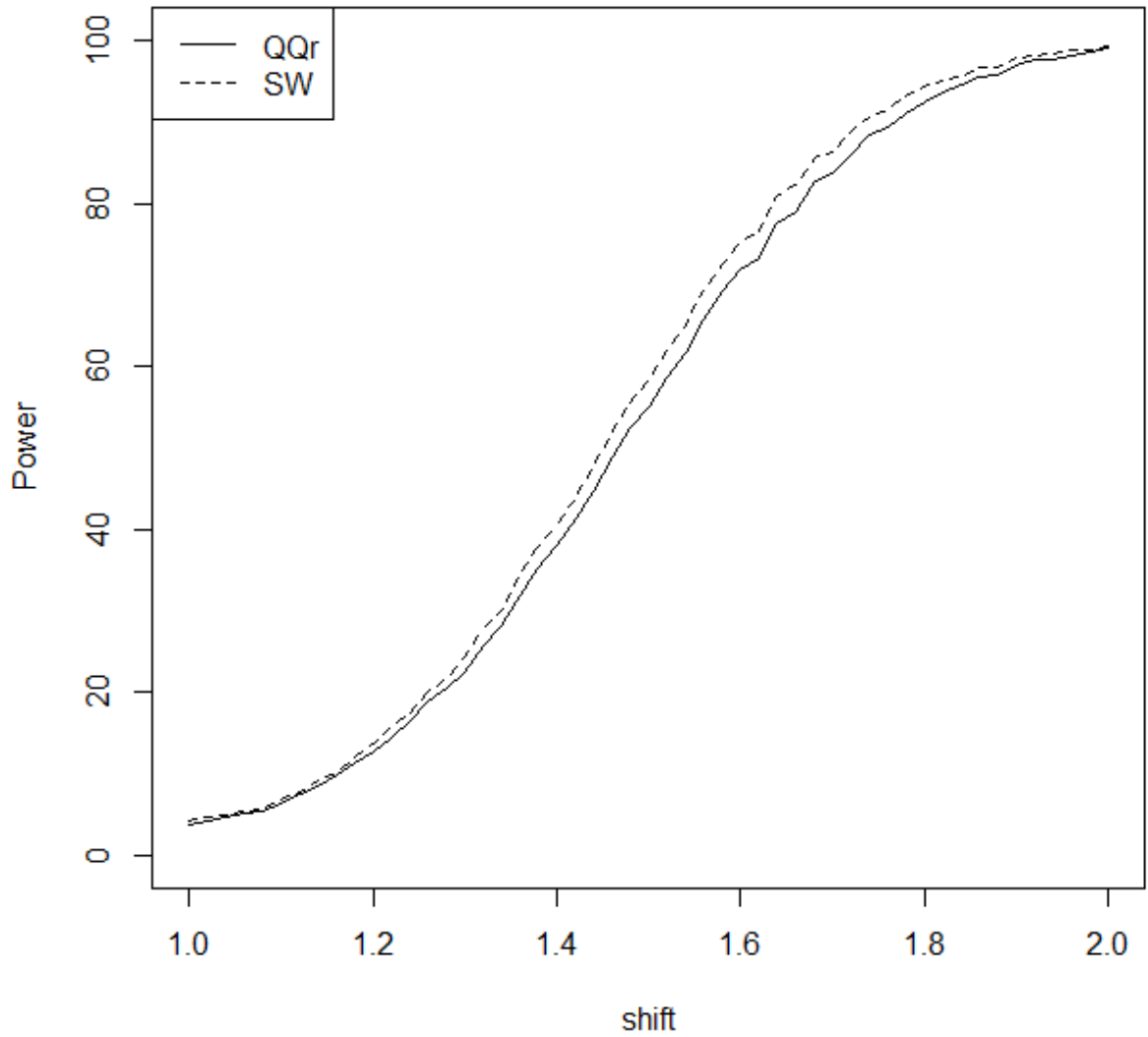

Figure G1. Comparing the power of QQr with Shapiro-Wilk

## Appendix H.  Fitting the t distribution

The framework can also be used to fit other location-scale-shape families.  For example the student t distribution is sometimes used as a model for data close to normal but with heavier tails.  In this setting, the measurand $X$ is modeled by

$$\frac{X-\mu}{\sigma} \sim t_\nu$$

Here $\mu$ is still the mean of $X$, but $\sigma$, a scale parameter, is no longer the standard deviation of $X$.  In the data modeling context, $\nu$ is an additional parameter that needs to be estimated, much as the BC power is.  To estimate $\nu$ from a QQ plot, for each trial value of $\nu$ calculate the t scores

$$T_i = F^{-1}[(i-0.5)/n, \nu)$$

where $F^{-1}$ is the inverse cumulative t.  Regressing the ordered $X$ on these $T$ then gives:

- A QQ correlation,
- A regression intercept,
- A regression slope.

The value maximizing the correlation is the estimate of $\nu$ and the corresponding intercept and slope provide the estimates of $\mu$ and $\sigma$

To illustrate the possibilities, a data set of size 120 with $\mu=20$ $\sigma=4$ and $n=5$ was drawn from the t distribution with 5 degrees of freedom using the R commands

```
set.seed(1093)
X <- 20 + 4*rt(120, 5)
```

The true upper reference limit is $20+4F^{-1}(0.975,5) = 30.28$. Figure H1 shows a normal QQ plot on the left. The data set fails the test for normality. Its sigmoid shape suggests the possibility of modeling with the t distribution; the result is shown on the right. The correlation is maximized at $n = 3.7$ whose QQr = 0.9944. At this stage we do not know whether that is high enough to pass a formal test of fit, but it would pass comfortably if it were produced from a normal QQ plot. The slope and intercept of the plot are 20.00 and 3.42 respectively, giving a reference limit of 29.80, very close to the true value.

Like the Box-Cox, the search for $\nu$ does not require expert statistical software, but could be implemented in a spreadsheet environment, though possibly with the search restricted to integer values of $\nu$.

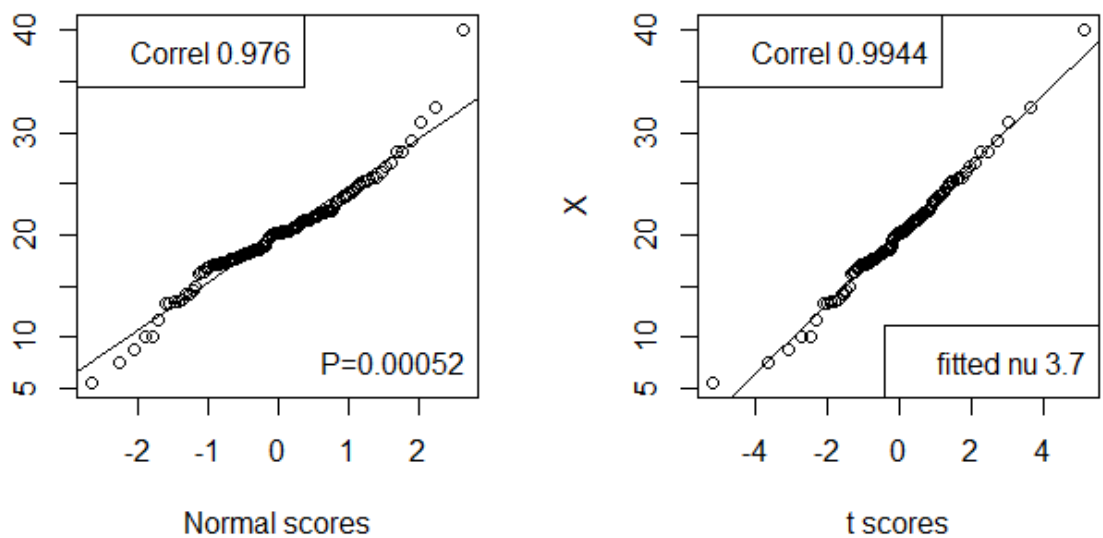

Figure H1. Fitting a t distribution

# Appendix I   Inverting the Box-Cox transformation

The Box-Cox transformation $Y = \dfrac{X^{\lambda} - 1}{\lambda}, \lambda \neq 0; \ = \ln(X), \ \lambda = 0$

aims to transform a measurand $X$ into a transform $Y$ that follows a $N(\mu, \sigma^2)$ distribution. The parameters $\mu, \sigma$ can then be estimated from the mean and standard deviation of the transform $Y$ in complete-data settings, or from the QQ plot with censored or winsorized data. Write $m$ and $s$ for these estimates. A reference range on the transformed scale can then be estimated as $m \pm z_\alpha s$, where $z_\alpha$ stands for the standard normal score for the desired coverage – in particular, for the default 95% central reference interval, $z_\alpha = 1.96$. Using the delta method, an approximate standard error for the 95% reference limits is then $e = s\sqrt{\dfrac{1}{n} + \dfrac{1.96^2}{2(n-1)}}$. The upper reference limit and its confidence interval are then $m+1.96s$ and $m+1.96s \pm z_\beta\, e$, where $z_\beta$ is the standard normal score for the desired confidence. In particular, for a 90% confidence interval, $z_\beta = 1.645$.

These three values – the estimate and its confidence interval – are on the transformed $Y$ scale. We might think to transform them back to the original scale by simply inverting the Box-Cox transformation to get $X = (\lambda Y + 1)^{1/\lambda}$ but doing so creates two difficulties:

- The back-transform is biased
- The confidence interval does not have the confidence level claimed.

The bias can be corrected, but for the sample sizes commonly seen in these studies is too small to cause concern. The shortfall in the confidence level though is a more serious concern. Linnet (1987) noted this and suggested that, for the particular sample size studied, it could be corrected by increasing the standard error by 25%. A more general approach is to use an equivalent effective sample size in the standard error calculation, making it

$$e = s\sqrt{\frac{1}{fn} + \frac{1.96^2}{2(fn-1)}}$$

where $f$ is some represents an equivalent sample fraction. This approach was tried for getting the 90% confidence interval for a 95% central reference limit for samples of size 40 (10) 100 (50) 200 (100) 1000, twelve values in all. For each sample size, an initial guess of $f=0.6$ was set. And then 50,000 N(0,1) samples were generated, exponentiated and analyzed by the Box-Cox. Motivated by the Robbins-Monro stochastic approximation algorithm, in the $i^{th}$ such sample, the estimate of the upper reference limit and its CI using the current $f$ were calculated, and scored as 1 if the CI covered the true upper reference limit and 0 if it did not. If it did not cover, suggesting that $f$ was too high, $f$ was multiplied by $(1-9/(i+20))$ for the next sample. If it did cover, suggesting that $f$ was too low, $f$ was multiplied by $(1+1/(i+20))$ for the next sample. This sequence of $f$ values covers the region of the true value, and converges to it, but to enhance performance, after ignoring an initial 200 warmup values, a logistic regression was fitted to the sequence of $f$ values and their 0-1 coverage, and the appropriate $f$ for 90% coverage computed. Figure I1, for $n=100$, illustrates the successive estimates $fn$ of the effective sample size and their coverage of the region of the reference limit. The horizontal dotted line is the refined value given by the logistic regression.

The resulting $f$ values are shown in Figure I2, along with the fitted curve

$$f = 0.68 - 5.09/n$$

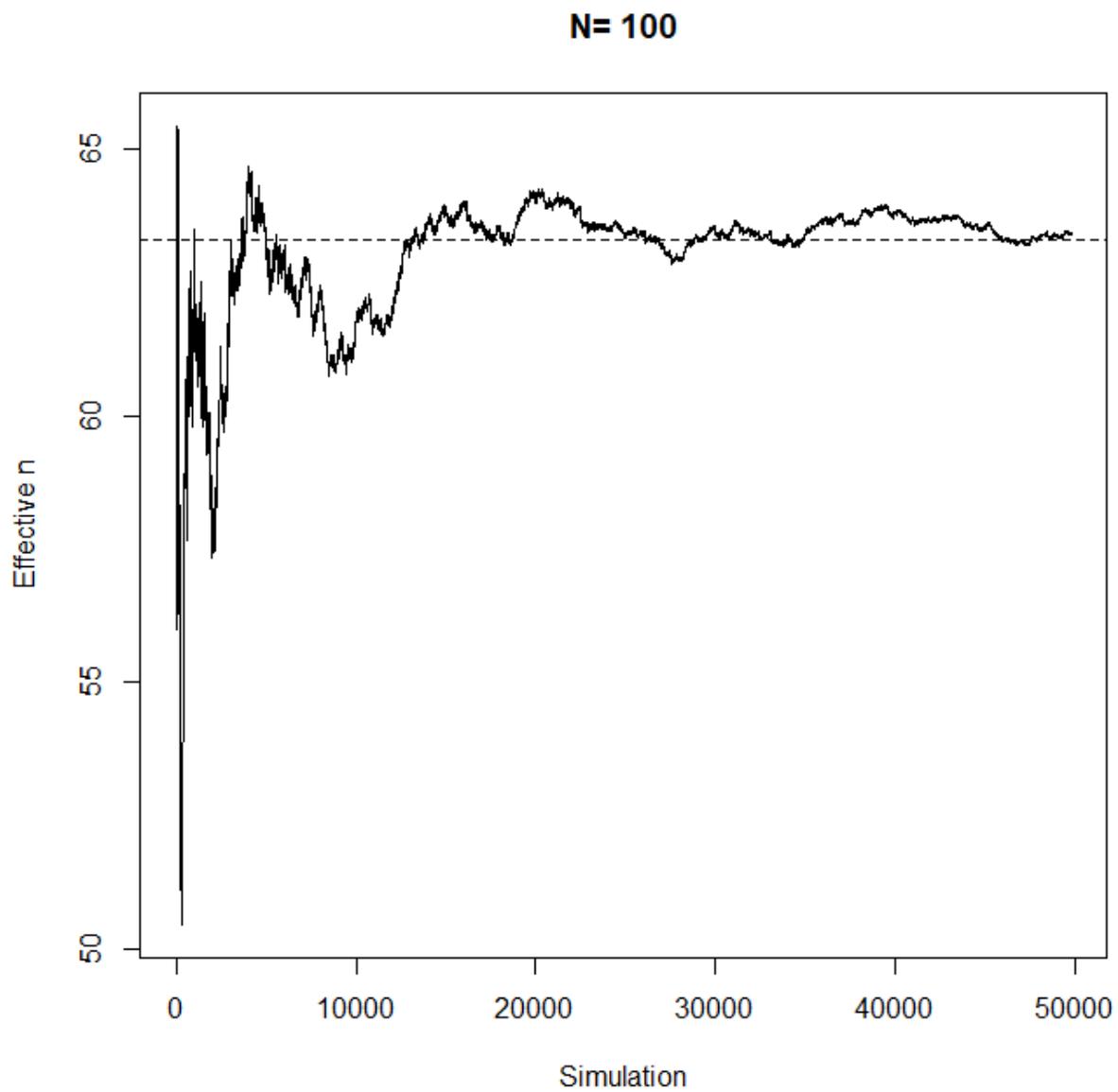

Figure I1.  Successive Robbins-Monro approximations and value from logistic regression

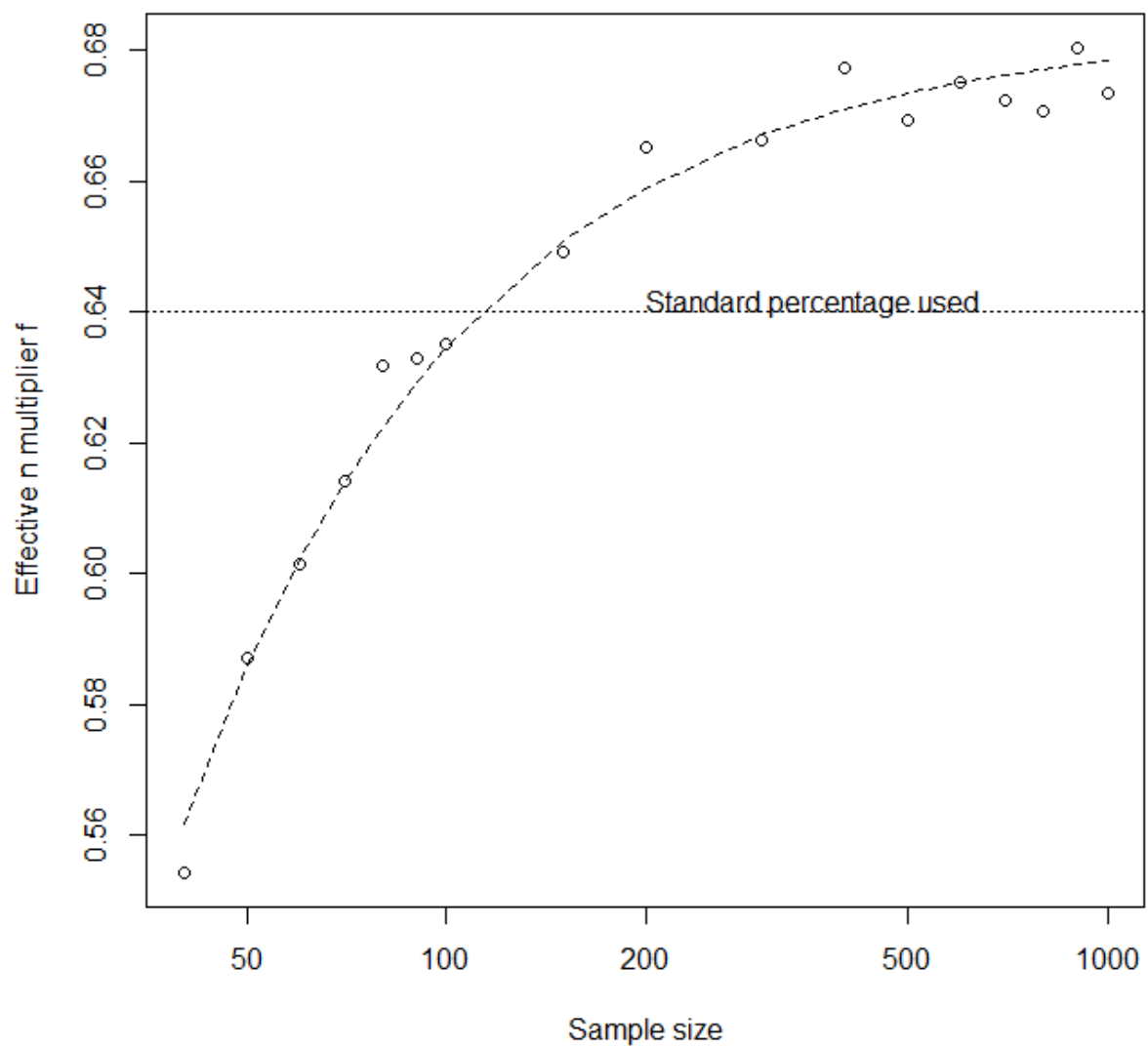

Figure I2.  Effective sample fraction as function of sample size

*References and Resources*